\documentstyle[epsfig]{article}
\begin{document}
\newcommand{\beq}{\begin{equation}}
\newcommand{\eeq}{\end{equation}}
\newcommand{\bea}{\begin{eqnarray}}
\newcommand{\eea}{\end{eqnarray}}
\newcommand{\cl}[1]{\begin{center} {#1} \end{center}}
\newcommand{\dr}{d^3r}
\newcommand{\dxp}{d^3x^{'}}
\newcommand{\dxpp}{d^3x^{''}}
\newcommand{\ddp}{\frac{d^3p}{(2 \pi)^3}}
\newcommand{\dpj}{\frac{d^3p_1}{(2 \pi)^3}}
\newcommand{\dpp}{\frac{d^3p^{'}}{(2 \pi)^3}}
\newcommand{\dpjp}{\frac{d^{3}p_{1}^{'}}{(2 \pi)^3}}
\newcommand{\dl}{\frac{d^3l}{(2 \pi)^3}}
\newcommand{\p}{{\bf p}}
\newcommand{\pj}{{\bf p_{1}}}
\newcommand{\pp}{{\bf p^{'}}}
\newcommand{\pjp}{{\bf p_{1}^{'}}}
\newcommand{\k}{{\bf k}}
\cl{\Large{{\bf Time development of a density perturbation in the
unstable nuclear matter}}} 
\vskip 1truecm

\cl{     P. Bo\.{z}ek \footnote{e-mail~: bozek@quark.ifj.edu.pl} }
 
\vskip 1truecm
\cl{ Yukawa Institute for Theoretical Physics, Kyoto University,
Kyoto 606-01 , Japan}
\cl{ and}
\cl{ Institute of Nuclear Physics, 31-342 Krak\'{o}w, Poland}
     

\begin{abstract}
We present the  solution of the time development of an unstable
initial density perturbation in the linearized Vlasov equation,
completing the previous analysis in the literature. The additional
contributions found are usually damped and can be neglected at large times
in the unstable region. The work clarifies also the problem of the
normalization of the solution with respect to the initial perturbation
of the density.
\end{abstract}

\vspace{0.5cm}


\vspace{0.5cm}

\noindent 
Keywords~: Vlasov equation, spinodal decomposition, unstable modes



With the development of the stochastic transport equations for nuclear 
collisions , i.e. the Boltzmann-Langevin equation, the description of
the multi fragmentation in intermediate energy nuclear collisions has
been
challenged \cite{c1,c2,chaos}.
 The spinodal decomposition mechanism of fragment formation is based
on the idea first proposed by Heiselberg et al. \cite{he}
that the nuclear matter at low density $\rho \simeq \rho_0/3$ is
unstable against density perturbation. In the linear regime, 
any perturbation with unstable
spatial momentum $k$ will develop in time   as $e^{\Gamma(k)t}$.
The fastest growing mode will dominate the linear response and
consequently, the typical fragments will have the size $1/k$ ,
determined
by the largest growth rate $\Gamma(k)$.
The importance of the nonlinearities of the Vlasov equations for the
fragment formation is still not settled \cite{chaos}. However, even
the linear response analysis presented so far are often incomplete.

It may seem that after many works dealing with the development of
the instabilities in the Fermi systems everything is understood in the
linear regime \cite{p1,p2,c2,c1}.  Usually, people study the proper modes
of the Fourier transformed linearized Vlasov equation~:
\beq
i({\bf k}{\bf v} - \omega) \delta f(\omega,k,p)
-i\frac{\partial U_k}{\partial \rho} {\bf k}{\bf v} \frac{\partial
n_0}{\partial \epsilon}\delta \rho(\omega,k) =0 \ ,
\eeq
where 
$\delta f(\omega,k,p)= \int dt \ d^3x \ \delta f(t,x,p) \ 
e^{i\omega t - i {\bf kx}}$ 
is the Fourier transformed in time and
space  perturbation of the equilibrium phase-space distribution $n_0$,
$U_k$ is the Fourier transformed density dependent mean field.
In the spinodal region the eigenvalue problem for
the linearized Vlasov equation has two
imaginary frequency solutions $\omega(k)=\pm i\Gamma(k)$ in certain
range
of $k < k_{max}$ limited by the range of the mean field potential.
The solution using the eigen-functions
 of the Vlasov equation is 
not a well posed   problem for the solution of 
 the time development of the initial
perturbation (or for the solution of the Vlasov equation with a noise
term \cite{c2}).
The analysis using   one sided Fourier
transform are more suitable and are a  well posed boundary value for
the Vlasov equation. 
In \cite{p1}, using the one-sided Fourier transform
 only the large time limit is explicitly
discussed. References \cite{c1,c2} discuss the singularities
corresponding to the non-collective modes. In the case of
discrete momentum these singularities correspond to the cut
singularity in the continuum momentum studied below.
In the framework studied in this letter the
calculations of the  contribution from the non-collective modes can be
easily done. In particular we can show that in non-exceptional cases
 of the initial perturbation the additional
contribution to the density development are damped. 
In the discrete formulation \cite{c2} it requires the use of a
large 
basis of  as complete as possible, since the damping can arise 
only from the beating in frequency of the sum of many contributions
 with real frequency.

Below, we shall calculate the evolution of the initial density
perturbation $\delta f(t=0,k,p)=g(k,p)$
in the linearized Vlasov equation in the unstable region.
Taking the one-sided Fourier transform ~:
\beq
\delta f(\omega,k,p) = \int_0^{\infty} dt f(t,k,p) e^{i \omega t}\ ,
\eeq
one obtains~:
\beq
i({\bf k}{\bf v} - \omega ) \delta f(\omega , k,p) - i
\frac{\partial U_k}{\partial \rho}{\bf k}{\bf v} \frac{\partial
n_0}{\partial \epsilon}\delta \rho(\omega,k)=g(k,p) \ .
\eeq
Dividing by ${\bf k}{\bf v} - \omega$ and integrating over $p$ an
equation for the density perturbation is obtained.
The time dependence of the density perturbation $\delta \rho$
 can be found using the inverse transform~:
\beq
\label{inv}
\delta \rho(t,k)= - i \int_{-\infty + i \sigma}^{\infty + i \sigma}
\frac{d \omega}{2
\pi}\frac{G(k,\omega)}{\epsilon(k,\omega)}e^{-i\omega t} \ ,
\eeq
with
\beq
G(k,\omega)=\int \frac{d^3p}{(2 \pi)^3}\frac{g(k,p)}{{\bf k}{\bf
v} - \omega} \ ,
\eeq
and 
\beq
\label{ep}
\epsilon(k,\omega)=1-\frac{\partial U_k}{\partial \rho} \int \frac{d^3 p}{(2
\pi)^3}\frac{{\bf kv}}{{\bf  kv}-\omega}\frac{\partial n_0}{\partial
\epsilon} \ ,
\eeq
the integration path in the inverse Fourier transform laying above
any singularity of the integrand. The integral can be calculated closing the
integration path in the lower half-plane.
The solutions of the dispersion relation~:
\beq
\label{disp}
\epsilon(k,\omega)|_k=0
\eeq
gives the poles of the integrand. The contribution to the integral from
the residues at these  poles 
gives the eigenfunction solution of the linearized Vlasov equation.
In the spinodal region, for the wave-vector ${\bf k}$ of 
an unstable mode 
 the dispersion relation (\ref{disp}) will have two solutions 
with imaginary frequencies.
The two frequencies give a damped and a growing component in the time
dependence of the density fluctuation~:
\beq
\label{pol}
\delta \rho_{pole}(t,k)=\sum_{\pm} \frac{-G(k,\pm i \Gamma(k))}{\partial
\epsilon(k,\omega)/
\partial \omega|_{\omega=\pm i\Gamma(k)}} e^{\pm\Gamma(k) t} \ .
\eeq
For large times the part with growing exponent dominates.
However, this solution is incomplete. For t=0 we do not recover the
initial condition~:
\beq
\delta \rho_{pole}(t=0,k) \neq \int\frac{d^3 p}{(2 \pi)^3}g(k,p) \ ,
\eeq
e.g. for
 \beq
\label{locd}
g(k,p)=(2 \pi)^3 g(k) \delta^3({\bf p}-{\bf p}_0)
\eeq
 at zero temperature
one gets for the overlap of the density disturbance with the
eigen-functions of the Vlasov equation~:
\beq
\frac{\delta \rho_{pole}(t=0,k)}{g(k)}=
\frac{-2 \Gamma(k)^2(\Gamma(k)^2+(kv_f)^2)}{(\Gamma(k)^2+(kv_f)^2(1+F_0(k)))
(\Gamma(k)^2+({\bf
kp}_0)^2)}\neq 1 \ ,
\eeq
where $F_0(k)=\frac{\partial U}{\partial \rho}\frac{m p_f}{2 \pi^2}$ 
is the Landau parameter $F_0$ of the Landau-Fermi liquid,  $p_f$
and $v_f$ denote the Fermi momentum and Fermi velocity respectively.
Also  generally, the sum of the  overlap with
the stable and with unstable eigen-functions is different from $1$~.
The original integral in the right hand of (\ref{inv}) has a
correct
limit at $t=0$~:
\beq
\delta \rho(t=0,k)= - i \int_{-\infty + i \sigma}^{\infty + i \sigma}
\frac{d \omega}{2
\pi}\frac{G(k,\omega)}{\epsilon(k,\omega)}=
\int\frac{d^3 p}{(2 \pi)^3}g(k,p)  \ ,
\eeq
as one can convince oneself calculating the residue of the integrand at
infinity.
The discrepancy between the two results arises from the contribution
of the cut in the integrand on the right hand side of (\ref{inv}) on the
real axis in frequency $\omega$.

 The contribution from the poles of the inverse susceptibility
$1/\epsilon(k,\omega)$ to the time development of the density
 is given in eq. (\ref{pol}). However, the two
poles are not the only singularities of the inverse susceptibility.
The singularity structure of the susceptibility on the real axis can be
read off after performing the angular  integration in (\ref{ep})~:
\beq
\epsilon(k,\omega)= 1 - F_0(k) \int_0^{\infty} x \ dx \ \Phi\big(
\frac{\omega}{x
k v_f}) 
\frac{\partial f_0(x)}{\partial x} \,
\eeq
where $\Phi(s)=1-\frac{s}{2}ln(\frac{s+1}{s-1})$ and
where in the rescaled variable $x$ the equilibrium distribution is
$f_0(x)= \frac{1}{exp({\epsilon_F(x^2-1)/T})+1}$~.
The logarithm in $\Phi(s)$ in the  the susceptibility has a cut on real axis~:
\bea
\label{epdi}
\epsilon(k,\omega\pm i \epsilon)& = & {\it Re}\  \epsilon(k,\omega) \pm
 i {\it Im}\  \epsilon(k,\omega) \nonumber \\
& = & 1 - F_0(k)\int_0^{\infty} x \ dx \ {\it Re} \Phi\big(\frac{
\omega}{x k v_f})
 \frac{\partial f_0(x)}{\partial x} \nonumber \\
 & &\pm i \frac{\pi F_0(k) \omega}{2 k v_f}
f_0\big(\frac{|\omega|}{kv_f}\big) \ .
\eea
Also the function $G(k,\omega)$ can only have singularities 
on the real axis.
Accordingly, the contribution of the cut to the density perturbation
is~:
\beq
\label{cut2}
\delta \rho_{cut}(t,k)=-i\int_{-\infty}^{\infty} \frac{d
\omega}{2 \pi} \bigg( \frac{G(k,\omega+i\epsilon)}
{\epsilon(k,\omega+i\epsilon)}- \frac{G(k,\omega-i\epsilon)}
{\epsilon(k,\omega-i\epsilon)}\bigg) e^{-i\omega t} \ .
\eeq
This is different from the cut singularity in
relativistic Vlasov equation where the
integration region is always finite.

Taking the perturbation well
localized in momentum (\ref{locd})
and using $\frac{1}{{\bf kv}_0-\omega \mp i\epsilon}=
P\frac{1}{{\bf kv}_0-\omega } \pm i \pi
\delta(\omega-{\bf kv}_0)$ we obtain~:
\bea
\delta \rho(t,k) & = & \frac{-1}{\pi}P\int_{-\infty}^{\infty}
\frac{d s}{s_0-s} \frac{{\it Im}\  \epsilon(k,s kv_f)}{
({\it Re}\  \epsilon(k,s kv_f))^2+({\it Im}\  \epsilon(k,s kv_f))^2}
e^{-i s kv_f t} \nonumber \\
& & +\frac{{\it Re}\  \epsilon(k,s_0 kv_f)}{
({\it Re}\  \epsilon(k,s_0 kv_f))^2+({\it Im}\  \epsilon(k,s_0 kv_f))^2}
e^{-i s_0 kv_f t} \ .
\eea
where $s_0=\frac{{\bf kv}_0}{kv_f}$~.
 Except for the k dependence of the Landau parameter
$F_0(k)$, the time scales appearing in the time dependence of the 
density $\delta \rho_{cut}(t,k)$ are proportional to $1/k$. It is the same
scaling as for the growth rates $\Gamma(k) \simeq k$ for constant
$F_0$~.
 In Fig. 1 we show the time dependence of $\frac{\delta
\rho_{cut}(t,k)}{g(k)}$ at $T=0$ and at 
$T=\epsilon_f/2$. The evolution of the density
$\delta 
\rho_{cut}(t,k)$  as given by the cut contribution is oscillating and
undamped for initial perturbation with definite momentum. The
phenomenon is similar
to  the Van Kampen modes in the relativistic Vlasov equation \cite{kin,flo}~.
Apparently, these modes have the overlap picked only on a finite
number of
eigenfunction in the  expansion in ref \cite{c2}.

Generally, the initial density disturbance will be a function of the
momentum $g(k,p)$. Although such a function can be written as a
superposition of $\delta$ functions in momentum, its behavior in time
will be different. The phenomenon is analogous the beating in frequency 
mentioned above. Let us study the evolution of a deformation of the
Fermi surface~:
\beq
\label{pp}
\delta f(t=0,k,p)=g(k) \frac{2 \pi^2 m}{{p_f} K_0 }\frac{ \partial n_0}{
\partial \epsilon} ,
\eeq
where $K_0=\int_{0}^{\infty} f_0(x) dx$.
The time development of the density disturbance is given by the sum
of the term from the poles of $1/\epsilon(k,\omega)$ and form the cut on the
real axis in frequency. The projection on the eigen-functions of the
density $G(\pm i \Gamma(k),k)$ will grow or will be damped exponentially
according to eq. (\ref{pol}). 
The overlap of the initial density perturbation with the eigen-functions of
the Vlasov equation as a function of the instability rate is shown in
Fig.2. 
 The overlap is greater than
$1$ with the limits $1$ and $2$ in the case of strong and weak
instability respectively.
This shows that the contribution from the cut cannot be
neglected at small times in order to insure the correct limit at $t=0$.
 The contribution from the cut can be
found from eq. (\ref{cut2}), with~:
\bea
G(\omega\pm i \epsilon,k)& = & - \frac{ g(k) }{2 K_0 kv_f} 
\int_0^{\infty} x \  dx \ ln \bigg| \frac{\omega+xkv_f}{\omega-xkv_f}
\bigg| \frac{\partial f_0(x)}{\partial x}
 \nonumber \\
& & \mp i \frac{ \pi g(k)}{2 K_0 kv_f}\int_{|\omega|/kv_f}^{\infty}
x  \frac{\partial f_0(x)}{\partial x} dx \ .
\eea
The result of the numerical integration is shown in Fig. 3. We see
that the contribution form the cut to the density distribution is
rapidly damped in time. As in the previous cases the time scales
involved in the time dependence of the density disturbance are
proportional to the wave-vector $k$, for constant $F_0$~.
For both cases of the initial distribution studied we obtain the
correct limit of the total density distribution at $t=0$, $\delta 
\rho_{pole}(t=0,k)+ \delta \rho_{cut}(t=0,k)=g(k)$~.
The finite temperature the curve is smooth, almost exponential, without the
oscillations found at zero temperature. These zero temperature
oscillations reflect the sharp limits of the integration in the
formula (\ref{cut2}), since at zero temperature the cut on the real axis
 is limited to the interval $|\omega| < k v_f$.

We have found that the description of the time development of an
initially unstable density perturbation can be described as a sum of two
terms. The term coming from the poles of the inverse 
susceptibility function, with
a growing and a damped mode. Unlike in the stable region the
discontinuity in the susceptibility function does not modify the result for
this contribution to the density evolution. In particular the position
of the zeros of the susceptibility function is not changed, in
contrast to the appearance of the Landau damping for the stable mode
of the Vlasov equation \cite{landau}.
 A second contribution was
found coming from the cut on the real axis in frequency for the
susceptibility function $\epsilon(k,\omega)$
 and for the one-sided Fourier transform of the density
perturbation $G(k,\omega)$. We have shown that this contribution is
bounded  in time, and
generally strongly damped. Only in the case of a singular perturbation
in momentum an oscillating solution for the contribution from the cut
was found. This contribution can be than neglected at large
times, where the unstable mode dominates. However, as it can be seen
in Fig.4,  it is
important at small times.
In particular the contribution from the cut integration
 to the density disturbance
is required in order to recover the initial density perturbation from the
solution of the Vlasov (\ref{inv}) equation at zero time. Indeed,
 the overlap of
the initial perturbation with the growing and damped mode is mostly
larger than $1$~. It is compensated by a negative contribution to the
density at zero time coming from the cut integration. It would be
incorrect to normalize the pole contribution to the density
disturbance at zero times $\delta \rho_{pole}(t=0,k)$ to $g(k)$~.
 The correct limit at
$t=0$
of the time development of the density perturbation originates from the 
following sum rule~:
\bea
\int \frac{d^3p}{(2 \pi)^3} f(p) & = & 
\sum_{\pm} \frac{-G(k,\pm i \Gamma(k))}{\partial
\epsilon(k,\omega)/
\partial \omega|_{\omega=\pm i\Gamma(k)}} \nonumber \\
& & -i\int_{-\infty}^{\infty} \frac{d
\omega}{2 \pi} \bigg( \frac{G(k,\omega+i\epsilon)}
{\epsilon(k,\omega+i\epsilon)}- \frac{G(k,\omega-i\epsilon)}
{\epsilon(k,\omega-i\epsilon)}\bigg)  \ ,
\eea
if
\beq
G(k,\omega)=\int \frac{d^3p}{(2 \pi)^3}\frac{f(p)}{{\bf k}{\bf
v} - \omega} \ .
\eeq
This sum rule corresponds to taking the complete basis for the
expansion of the initial perturbation in ref \cite{c2}.

 The results presented in this paper,
explain how the initial density disturbance gets additional growing
term corresponding to the rapid decay of the negative contribution from
the cut integration.
This effect is the most pronounced at small values of the growth rate
$\Gamma/k$, where the overlap of the density perturbation with the
eigen-functions  is larger than $1$, although finite. It modifies
the result of a simple growth of unstable eigen-functions of the Vlasov
equation by the value of the overlap at the initial time. This
enhances the initial perturbation for the wave-vectors corresponding to
small growth rate $\Gamma/k$. It means that that the prediction that
the most unstable modes would be preferably excited would be weakened,
since the overlap with unstable eigenfunction is close to $1/2$ for
large $\Gamma/k$. The contribution form the cut integration can be
neglected in this case. On the other hand,
 it gives a negative and strongly damped
contribution at small $\Gamma/k$, which results in a faster increase of
the density disturbance at small times. At large times, this effect
is taken into account as a normalization of the overlap of the initial
density perturbation	 with the unstable
mode different form $1/2$~.

We have presented the formula and figures in a general way,
 using  variables scaled
by $1/kv_f$. It shows that, except for the $k$ dependence of the Landau
parameter $F_0$, the time scales of the oscillation or decay of the 
contribution to the density disturbance coming form the cut
integration  scale with $1/kv_f$. Similarly as the growth rate
$\Gamma(k)\simeq kv_f$. This means that the decay times of $\delta
\rho_{cut}$ go to infinity as $k \rightarrow 0$.

\vspace{5mm}

\noindent
{\bf Acknowledgments} \\
The author wishes to thank  for the hospitality extended to him
by the YITP.
 

\newpage

 
\begin{figure}
\begin{center}
\epsfig{file=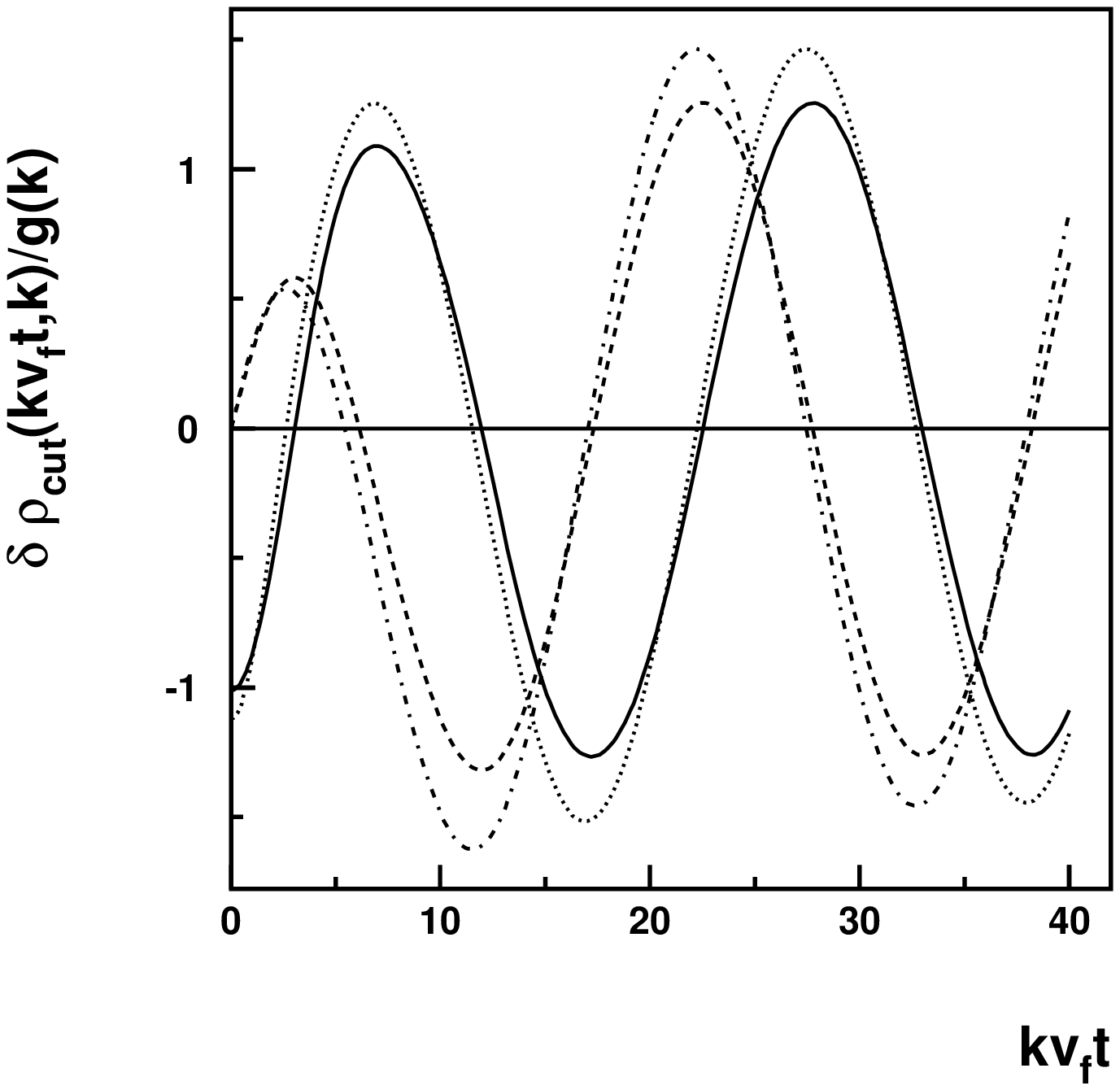,width=0.8\textwidth}
\vspace{0.5cm}
\end{center}
\end{figure}

{\bf Fig.1} \\
The time development of the initial density perturbation singular in
momentum (\ref{locd})
 ($F_0=-1.5$, ${\bf kv}_0=0.5 kv_f$).
The solid line and the dashed line represent respectively 
the real and imaginary
part of the density perturbation at zero temperature.
The dotted line and the dashed-dotted line represent respectively 
the real and imaginary
part of the density perturbation at $T=\epsilon_f/2$.

\newpage
\begin{figure}
\begin{center}
\epsfig{file=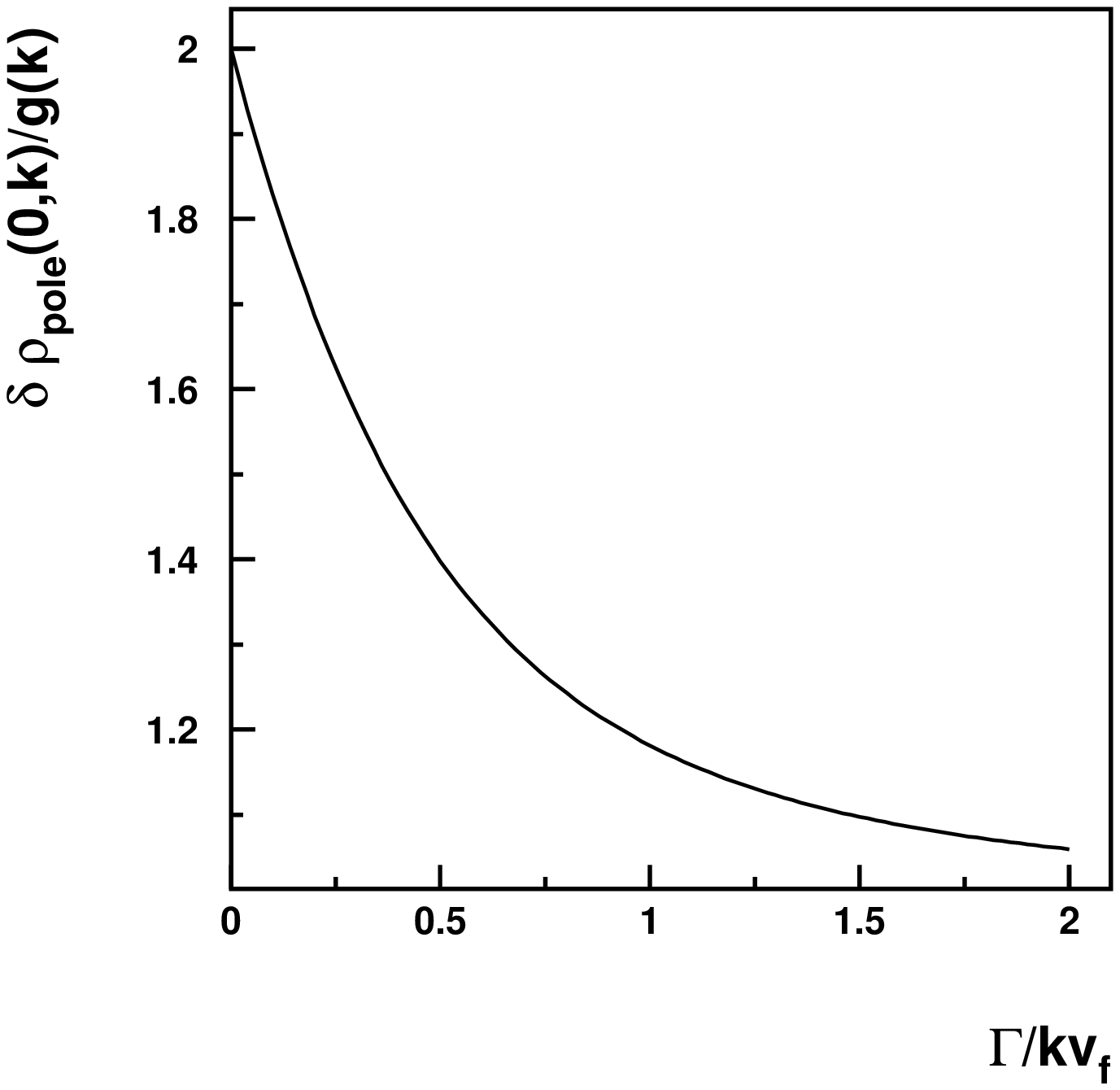,width=0.8\textwidth}
\vspace{0.5cm}
\end{center}
\end{figure}

{\bf Fig. 2} \\
The sum of the overlap of the density perturbation  (\ref{pp})
with the eigen-modes of the Vlasov equation as function of the 
growth (damping) rate $\Gamma$ at zero temperature. Very little change
occurs at non-zero temperature.

\newpage
\begin{figure}
\begin{center}
\epsfig{file=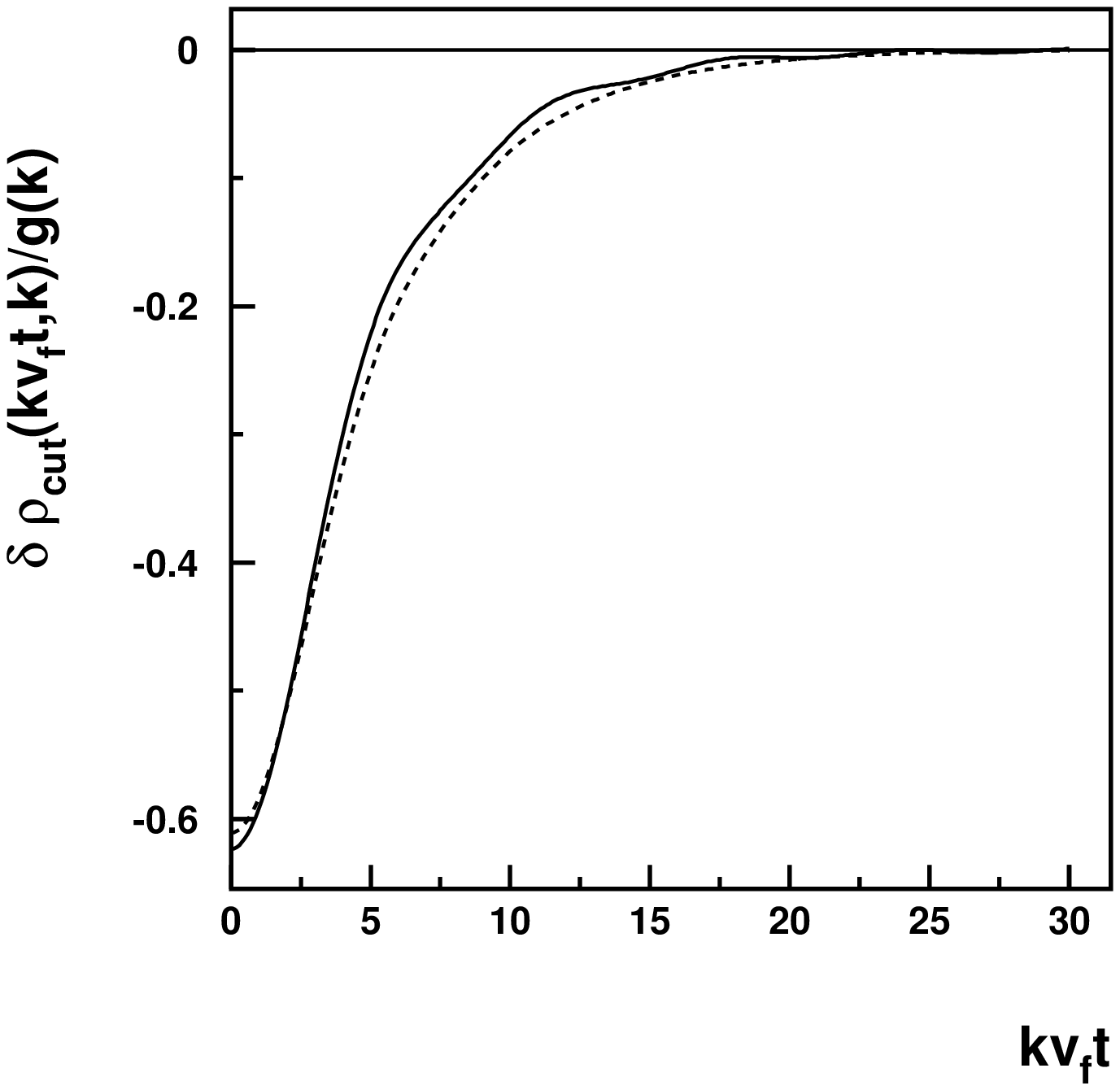,width=0.8\textwidth}
\vspace{0.5cm} 
\end{center}
\end{figure}

{\bf Fig. 3} \\
The time development of the initial density perturbation(\ref{pp})
 ($F_0=-1.5$).
The solid line and the dashed line represent the result at zero
temperature
and at
$T=\epsilon_f/4$ respectively.

\newpage
\begin{figure}
\begin{center}
\epsfig{file=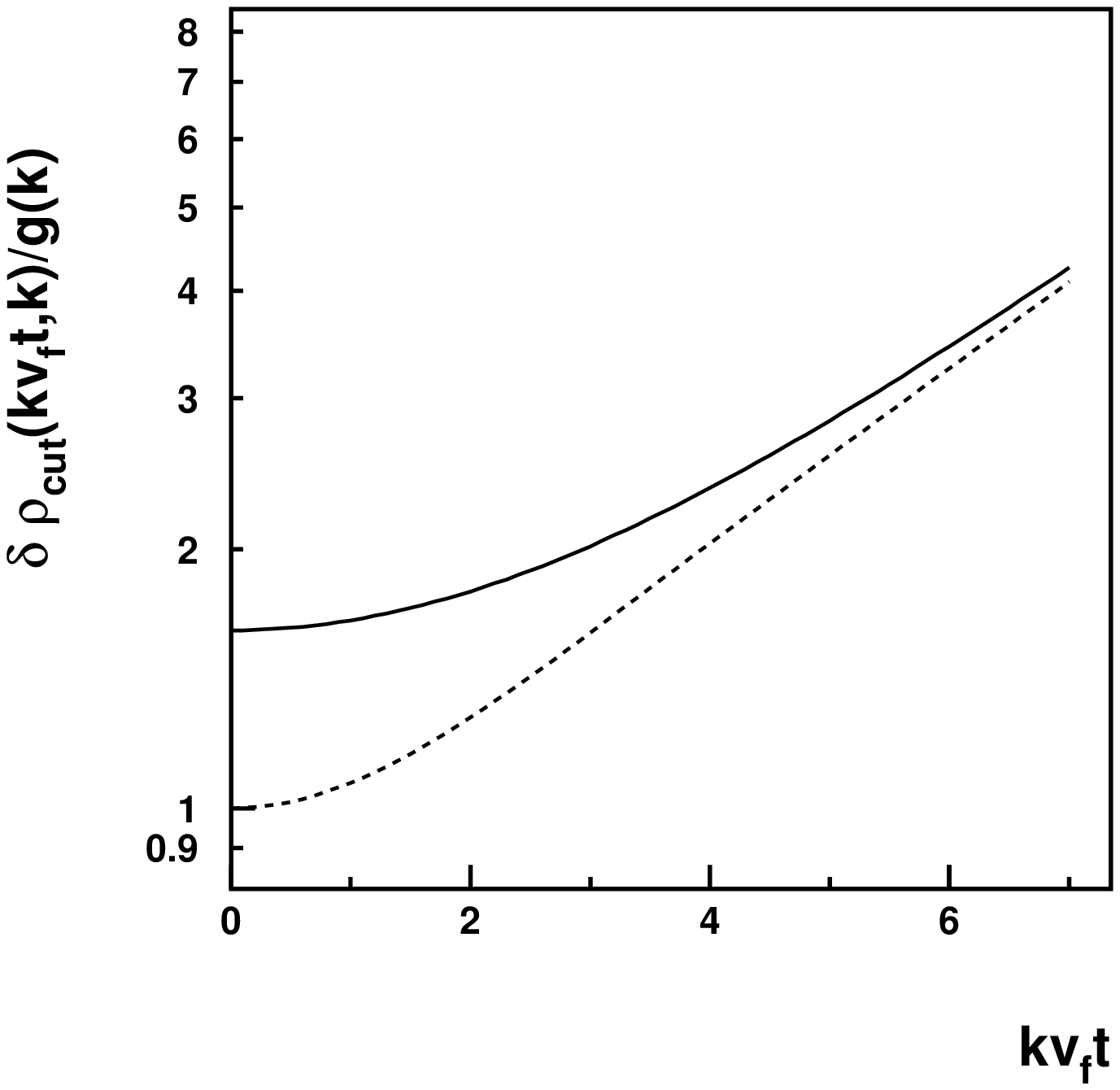,width=0.8\textwidth}
\vspace{0.5cm} 
\end{center}
\end{figure}

{\bf Fig. 4} \\
The time development of the initial density perturbation 
 (\ref{pp}) at zero temperature
 ($F_0=-1.5$).
The solid line represents the  pole contribution.
The dashed line represents total density perturbation, i.e.
 the sum of the pole and the cut contribution.

\end{document}